\documentclass[12pt]{article}

\usepackage[small]{caption}
\usepackage{epsfig} 

\newenvironment{myitemize}{\begin{list}{${\mbox{\boldmath $-$}}$}{\topsep=2mm \parsep=2mm \itemsep=0mm \leftmargin=8mm}}{\end{list}}

\title{
  \vspace*{-2cm} 
  \begin{flushleft} {\small\tt Talk presented at the SCCS 99, Saint-Malo (France), Sept 4--11, 1999} \end{flushleft}
  \vspace{5mm} 
  \large\bf
  Dynamics of the 2D two-component plasma \\
  near the Kosterlitz-Thouless transition
  \vspace{-2mm} 
}

\author{
  \normalsize Dierk Bormann \vspace{-2mm} \\ 
  \small ({\sl Universit\"at Augsburg, Germany}) \\
  \normalsize Hans Beck and Oliver Gallus \vspace{-2mm} \\
  \small ({\sl Universit\'e de Neuch\^atel, Switzerland}) \\
  \normalsize Massimiliano Capezzali \vspace{-2mm} \\
  \small ({\sl Queens University, Canada}) 
  \vspace*{-3mm}
}

\date{}

\begin{document}

\maketitle
\vspace{-5mm}

\begin{itemize}
\item[]
{\small\noindent
{\bf Abstract.}
We study the dynamics of a classical, two-component plasma in two dimensions, in the vicinity of the Kosterlitz-Thouless (KT) transition where the system passes from a dielectric low-temperature phase (consisting of bound pairs) to a conducting phase. 
We use two ``complementary'' analytical approaches and compare to simulations. 
The conventional, ``intuitive'' approach is built on the KT picture of independently relaxing, bound pairs. 
A more formal approach, working with Mori projected dynamic correlation functions, avoids to assume the pair picture from the start. 
We discuss successes and failures of both approaches, and suggest a way to combine the advantages of both.
}
\end{itemize}

\vspace{-3mm}

\vspace{5mm}\noindent{\bf 1. INTRODUCTION}\par\vspace{3mm}\noindent%
%
The two-component plasma in two dimensions (2D) is the generic model for topological excitations in 2D systems with a U(1) order parameter symmetry, usually called {\em vortices} \cite{Minnhagen87}.
Prominent examples of such systems are the ``2D superfluids'', superconducting or $^4$He films and Josephson junction arrays (JJA's), others are the $XY$ model of 2D planar magnets, and --- with some modifications --- 2D melting \cite{Nelson83}.
The 2D plasma undergoes the well-known Kosterlitz-Thouless (KT) transition at a finite temperature $T_{\rm\scriptscriptstyle KT}$.

The dynamic behavior close to this transition is of both principal and practical interest.
It is conveniently described by a frequency dependent, complex dielectric function $\epsilon(\omega)$, 
related to the dynamic correlation function $\Phi_\rho({\mbox{\boldmath $k$}},\omega)$ of the charge density $\rho({\mbox{\boldmath $r$}},t)$ of the plasma:
\begin{eqnarray}
  \frac{1}{\epsilon(\omega)} 
  & = & 1 - \left. \frac{2\pi}{k_{\rm\scriptscriptstyle B} T\, k^2} \left[ q^2 \bar{n} S_\rho({\mbox{\boldmath $k$}}) + {\rm i}\omega \Phi_\rho({\mbox{\boldmath $k$}},\omega) \right] \right|_{{\mbox{\boldmath ${\scriptstyle k}$}} \to {\mbox{\boldmath ${\scriptstyle 0}$}}}
\ ,\label{eq:i2}\\
  \Phi_\rho({\mbox{\boldmath $k$}},\omega) 
  & = & \int\limits_{0}^{\infty} {\rm d} t \,{\rm e}^{{\rm i}\omega t} \int\limits_{}^{} {\rm d}^2{\mbox{\boldmath $r$}} \,{\rm e}^{-{\rm i} {\mbox{\boldmath ${\scriptstyle k}$}}\cdot{\mbox{\boldmath ${\scriptstyle r}$}}} \bigl\langle \rho({\mbox{\boldmath $r$}},t) \rho({\mbox{\boldmath $0$}},0) \bigr\rangle 
\ .\label{eq:i3}\end{eqnarray}
Here $\bar{n}$ is the mean number density of the particles with charge $\pm q$, and $S_\rho({\mbox{\boldmath $k$}}) = \Phi_\rho({\mbox{\boldmath $k$}},t\mbox{$=$}0)/q^2\bar{n}$ is the static charge structure factor, determining the zero frequency limit of $\epsilon(\omega)$.
As usual, $\epsilon(\omega)$ is related to the dynamic conductivity $\sigma(\omega)$ by $\epsilon(\omega) = 1 - 2\pi \sigma(\omega)/{\rm i}\omega$.

Recent numerical simulations by Jonsson and Minnhagen (JM) \cite{Jonsson+Minnhagen97a} of the 2D $XY$ model with Ginzburg-Landau dynamics have shed new light onto the critical dynamics of the associated vortex plasma. 
Using an appropriate definition for the vortex density $\rho({\mbox{\boldmath $r$}})$, 
JM have computed $\Phi_\rho$ and extracted the complex dielectric function $\epsilon(\omega)$ using (\ref{eq:i2}). 
It shows the following, interesting features:

\begin{myitemize}

\item
$1/\epsilon(\omega)$ approximately has a scaling form with a characteristic frequency $\omega_{\rm ch}$ going to zero as $T \to T_{\rm\scriptscriptstyle KT}$ (``critical slowing down'') from both sides.

\item
For low frequencies, $\,{\rm Re}[1/\epsilon(\omega) - 1/\epsilon(0)] \propto \omega$ 
over a relatively large $\omega$ interval, contrary to a simple ``Drude'' behavior ($\propto \omega^2$).
This corresponds to a non-analyticity $\sigma(\omega) \propto \ln|\omega|$ at low $\omega$.

\end{myitemize}

\noindent
JM obtain very good fits of their data by a phenomenological expression proposed earlier \cite{Minnhagen87} by Minnhagen.
Experiments on JJA's \cite{Theron+93} seem indeed to confirm the non-analytic frequency dependence of $\epsilon(\omega)$, and various theoretical attempts have been made \cite{anom_dyn} in order to explain this finding.
The critical slowing down has, to our knowledge, not been theoretically addressed so far.

\vspace{5mm}\noindent{\bf 2. AHNS PAIR RELAXATION PICTURE}\par\vspace{3mm}\noindent%
%


\begin{figure}[t]{
\vspace*{0mm}\centerline{\hspace{-7mm}
\epsfysize=69mm\epsffile{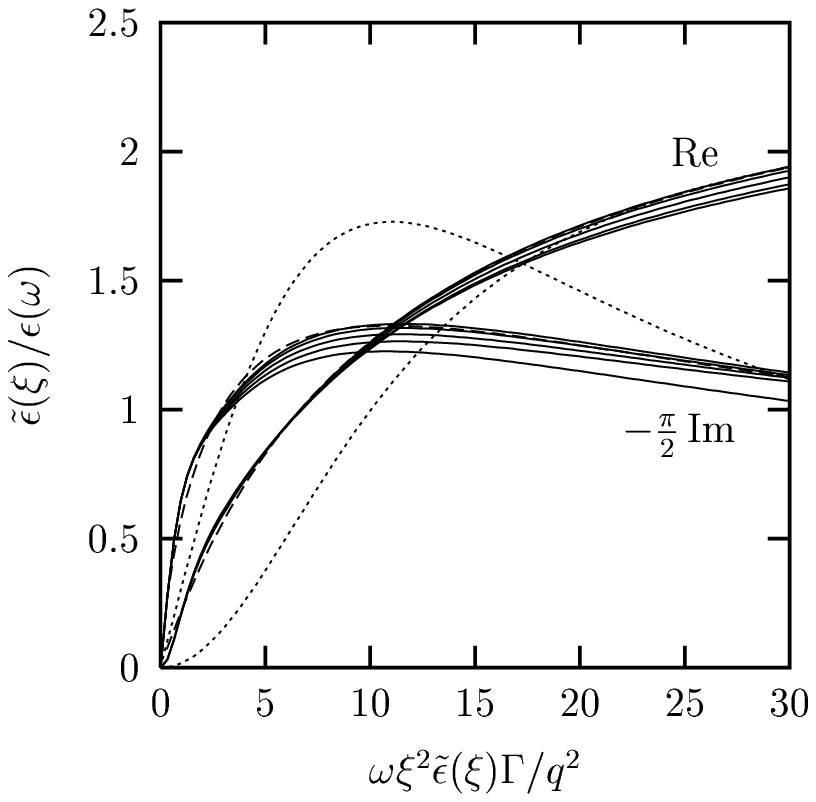}\hspace{5mm}
\epsfysize=69mm\epsffile{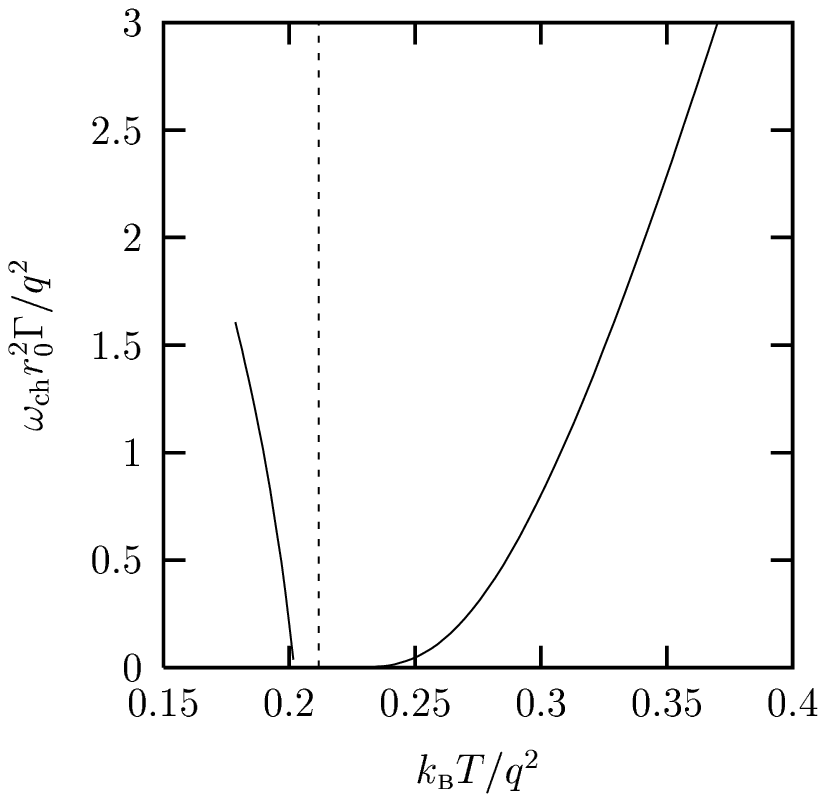}\hspace{0mm}
}\vspace*{-3mm}
}\caption{
Left: Real and imaginary parts of $\tilde\epsilon(\xi)/\epsilon(\omega)$ plotted against a scaled frequency for a series of different temperatures above the KT transition 
(full curves, $t_0 = -0.18, -0.15, -0.09, 0, 0.09$, and $\xi/r_0 = 6 \times 10^{23}, 9134, 63, 8.7, 3.7$ from above to below).
The scaling function is almost indistinguishable from the simple phenomenology of JM (dashed).
For comparison, a simple Drude form is also shown (dotted).
\mbox{Right: The characteristic frequency} $\omega_{\rm ch}$ defined in the text as a function of temperature $T$, above as well as below $T_{\rm\scriptscriptstyle KT}$ (= vertical dashed line).
}\label{fig:scal}\end{figure}


\noindent
The ``standard'' dynamic theory of the KT transition has been developed in 1978 by Ambegaokar {\sl et al.}\ (AHNS) \cite{AHNS},
who built up the dielectric function $\epsilon(\omega)$ additively from contributions of pairs of any size and of free particles.
We use a slight modification of this idea here, defining by
\begin{equation}
  \frac{1}{\epsilon_{\rm b}(\omega)} = 1 + \int\limits_{r_0}^{\xi} {\rm d} r \frac{{\rm d} [1/\tilde\epsilon(r)]}{{\rm d} r} \frac{1}{1 - {\rm i}\omega \tau(r)}
\ ,\quad
  \tau(r) \approx \frac{r^2 \tilde\epsilon(r) \Gamma}{3.5\, q^2}
\ \label{eq:a3}\end{equation}
a dielectric function of bound pairs, and setting $\epsilon(\omega) = \epsilon_{\rm b}(\omega) - 2\pi q^2 n_{\rm f} / {\rm i}\omega \Gamma$.
The scale dependent dielectric constant $\tilde\epsilon(r)$ is determined by the KT flow equations \cite{Minnhagen87, Nelson83},
and the factor $1/[1 - {\rm i}\omega \tau(r)]$ in the integral describes the relaxation of individual pairs of size $r$.
$r_0$ is a minimum length of the order of the vortex core size, $\xi$ is the KT screening length,
and $n_{\rm f} \approx \xi^{-2}$ the density of free particles.
To determine $\tilde\epsilon(r)$, we use Minnhagen's \cite{Minnhagen87} version of the KT flow equations for the scale dependent ``reduced temperature'' $t(r) = 1-q^2/4\tilde\epsilon(r)k_{\rm\scriptscriptstyle B} T$ and fugacity $y(r)$. 
Their solutions are known analytically, in the regions $T<T_{\rm\scriptscriptstyle KT}$ and $T>T_{\rm\scriptscriptstyle KT}$ separately \cite{Minnhagen87}.
The integral (\ref{eq:a3}) is done numerically. 

We find that the frequency at which $\,{\rm Im}[1/\epsilon(\omega)]$ has its maximum is more or less temperature independent below $T_{\rm\scriptscriptstyle KT}$, whereas above it varies rapidly like $\xi(T)^{-2}$.
Both above and sufficiently far below $T_{\rm\scriptscriptstyle KT}$, the low-$\omega$ behavior of $\,{\rm Re}[1/\epsilon]$ is closer to linear than a simple Drude form.
However, above $T_{\rm\scriptscriptstyle KT}$ there finally is a crossover to ``Drude'' $\omega^2$ behavior as $\omega \to 0$, and below $T_{\rm\scriptscriptstyle KT}$ both real and imaginary part of $1/\epsilon(\omega) - 1/\epsilon(0)$ behave as $\omega^\alpha$ at low $\omega$, with an exponent $\alpha = \frac{q^2}{2 \epsilon(0) k_{\rm\scriptscriptstyle B} T} - 2$ which goes to zero as $T \to T_{\rm\scriptscriptstyle KT}$.
If normalized by a factor of $\tilde\epsilon(\xi)$, our data above $T_{\rm\scriptscriptstyle KT}$ display an almost perfect scaling behavior with a characteristic frequency $\omega_{\rm ch}(\xi) \approx 10 q^2 / \xi^2 \tilde\epsilon(\xi) \Gamma$, over an extremely wide range in $\xi$ (Fig.\ \ref{fig:scal}\,L).
We find no comparable scaling below $T_{\rm\scriptscriptstyle KT}$, but if $\omega_{\rm ch}$ is defined here by the condition $\,{\rm Re}[1/\epsilon(\omega_{\rm ch}) - 1/\epsilon(0)] = - \frac{\pi}{2} \,{\rm Im}[1/\epsilon(\omega_{\rm ch})]$ (as done by JM), it decreases rapidly as $T \to T_{\rm\scriptscriptstyle KT}$ from {\em both} sides.
In Fig.\ \ref{fig:scal}\,R we have plotted the corresponding temperature dependences of $\omega_{\rm ch}$ both above and below $T_{\rm\scriptscriptstyle KT}$, which is in good qualitative agreement with Fig.\ 6 of JM \cite{Jonsson+Minnhagen97a}.


\begin{figure}[t]{
\vspace*{0mm}\centerline{\hspace{-5mm}
\epsfysize=69mm\epsffile{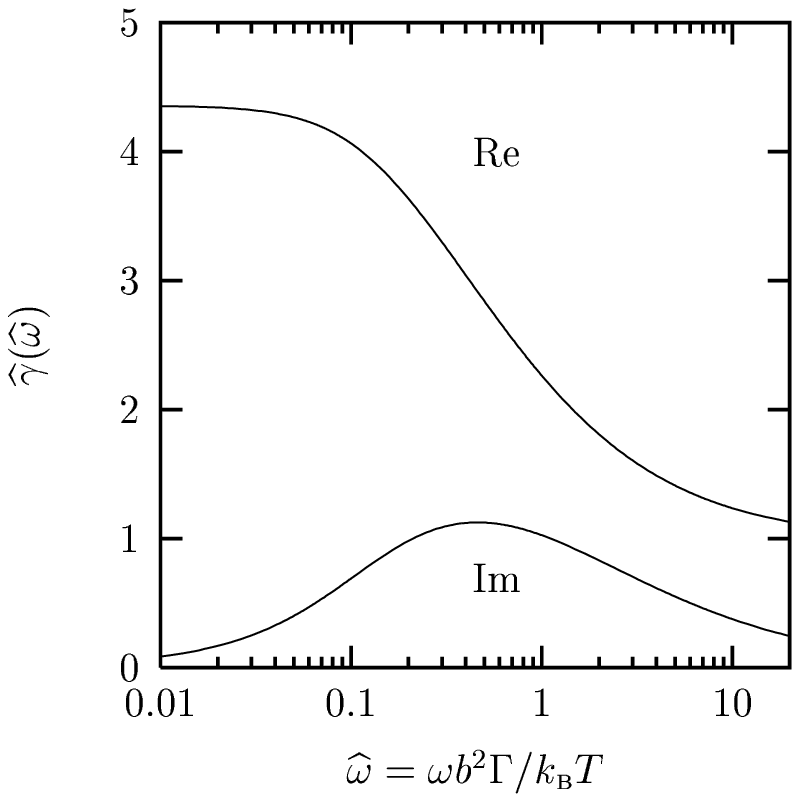}\hspace{15mm}
\epsfysize=69mm\epsffile{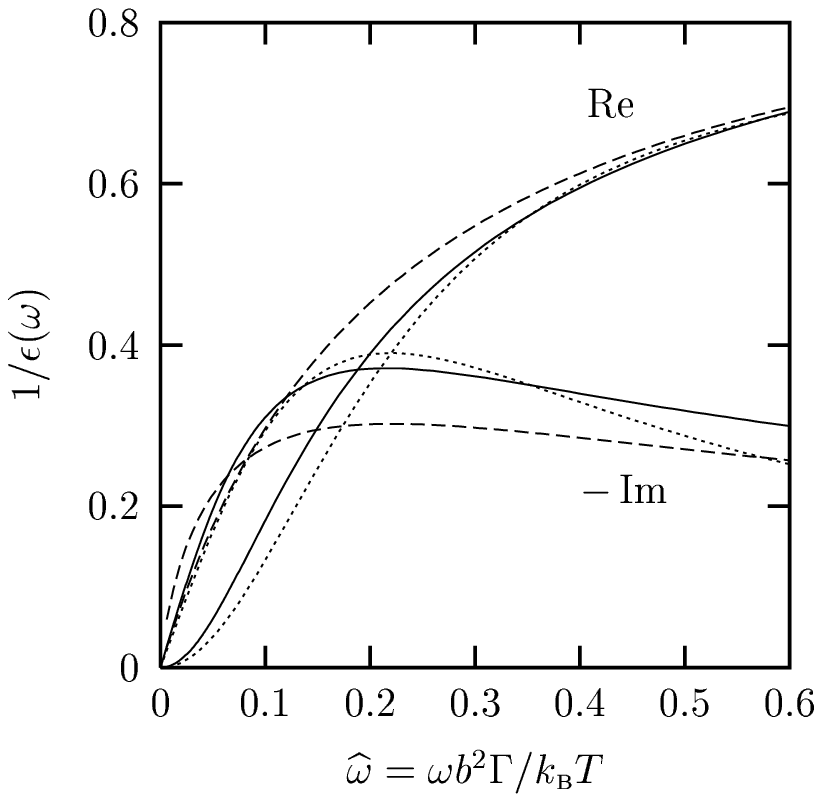}\hspace{0mm}
}\vspace*{-3mm}
}\caption{
Left: Real and imaginary parts of the reduced charge friction function $\widehat\gamma(\widehat\omega)$, calculated from Eq.\ (\protect\ref{eq:b8}) with $A=2$.
Right: The resulting $1/\epsilon(\omega)$ (full curves) compared to the phenomenology of JM (dashed) and to a Drude Form (dotted), with maxima of $-\,{\rm Im}$ at the same frequency ($\widehat\omega \approx 0.22$) and amplitudes adjusted.
}\label{fig:mori}\end{figure}



\vspace{5mm}\noindent{\bf 3. CALCULATIONS USING MORI'S TECHNIQUE}\par\vspace{3mm}\noindent%
%

\noindent
Mori's technique \cite{Hansen+McDonald86} for evaluating dynamic correlation functions yields for the charge density correlator at small wave numbers $k$:
\begin{equation}
  \Phi_\rho({\mbox{\boldmath $k$}},\omega) \sim \frac{q^2 \bar{n} S_\rho({\mbox{\boldmath $k$}})}{-{\rm i}\omega + \frac{k_{\rm\scriptscriptstyle B} T \, k^2}{S_\rho({\mbox{\boldmath ${\scriptstyle k}$}}) \gamma_\rho(\omega)}}
\ .\label{eq:b1}\end{equation}
The ``memory function'' (or ``random force'' correlation function) $\gamma_\rho(\omega)$ obeys
\begin{equation}
  \gamma_\rho(\omega) \approx \Gamma + \frac{q^2}{2 k_{\rm\scriptscriptstyle B} T \, \bar{n} V^3}
  \sum\limits_{{\mbox{\boldmath ${\scriptstyle k}$}} \, {\mbox{\boldmath ${\scriptstyle l}$}}}^{} {\mbox{\boldmath $k$}} \mbox{$\cdot$} {\mbox{\boldmath $l$}} \, U_{\mbox{\boldmath ${\scriptstyle k}$}} U_{-{\mbox{\boldmath ${\scriptstyle l}$}}} 
  \int\limits_{0}^{\infty} {\rm d} t \,{\rm e}^{{\rm i}\omega t}
  \bigl\langle \delta n_{-{\mbox{\boldmath ${\scriptstyle k}$}}}(t) \rho_{\mbox{\boldmath ${\scriptstyle k}$}}(t) \delta n_{\mbox{\boldmath ${\scriptstyle l}$}}(0) \rho_{-{\mbox{\boldmath ${\scriptstyle l}$}}}(0) \bigr\rangle
\ .\label{eq:b2}\end{equation}
where $\delta n_{\mbox{\boldmath ${\scriptstyle k}$}}(t)$ denotes the deviation of the local number density from its mean value, and $V$ is the system volume.
$\gamma_\rho(\omega)$ generalizes the bare friction constant $\Gamma$, adding a contribution due to the particle interaction.

In the spirit of current practice in the theory of liquid dynamics \cite{Hansen+McDonald86} we factorize, as a first step, the combined correlator into a product of a number and a charge correlation function:
\begin{equation}
  \bigl\langle \delta n_{-{\mbox{\boldmath ${\scriptstyle k}$}}}(t) \rho_{\mbox{\boldmath ${\scriptstyle k}$}}(t) \delta n_{\mbox{\boldmath ${\scriptstyle l}$}}(0) \rho_{-{\mbox{\boldmath ${\scriptstyle l}$}}}(0) \bigr\rangle
  \approx V^2 \delta_{{\mbox{\boldmath ${\scriptstyle k}$}}{\mbox{\boldmath ${\scriptstyle l}$}}} \Phi_n(-{\mbox{\boldmath $k$}},t) \Phi_\rho({\mbox{\boldmath $k$}},t)
\ .\label{eq:b3}\end{equation}
Using 
$\Phi_n({\mbox{\boldmath $k$}},\omega) \approx \bar{n} / ( -{\rm i}\omega + \frac{k_{\rm\scriptscriptstyle B} T \, k^2}{\Gamma} )$
and 
$S_\rho({\mbox{\boldmath $k$}}) \approx k^2/(k^2 + b^{-2})$
(the length $b$ is a measure of the interparticle distance)
we obtain the implicit equation
\begin{equation}
  \widehat\gamma(\widehat\omega) \approx 1 + \frac{A}{1 + {\rm i}\widehat\omega} \ln\frac{1 + \widehat\gamma(\widehat\omega)}{1 - {\rm i}\widehat\omega \widehat\gamma(\widehat\omega)}
\ ,\label{eq:b8}\end{equation}
determining the function $\widehat\gamma(\widehat\omega) = \gamma_\rho(\omega)/\Gamma$ which depends on $\widehat\omega = \omega/\omega_{\rm ch,1}$.
The characteristic frequency $\omega_{\rm ch,1} = k_{\rm\scriptscriptstyle B} T / b^2 \Gamma$ reflects the spatial density of particles, and $A = \frac{\pi q^4 \bar{n} b^2}{2(k_{\rm\scriptscriptstyle B} T)^2}$ is a dimensionless parameter of order unity.

Fig.\ \ref{fig:mori}\,L shows real and imaginary parts of the solution of (\ref{eq:b8}) above $T_{\rm\scriptscriptstyle KT}$, where $\sigma(\omega) \propto \gamma_\rho(\omega)^{-1}$. 
$\,{\rm Re}\widehat\gamma$ shows the expected (see Sec.\ 1) logarithmic $\omega$ dependence in some region, but has a finite limit for $\omega \to 0$.
As seen in Fig.\ \ref{fig:mori}\,R, our mode decoupling approach yields a dielectric function somewhat closer to the phenomenology of JM than a pure Drude form.
However, the critical slowing down is missing, since the factorization approximation (\ref{eq:b3}) underestimates effects of charge binding into pairs on length scales $< \xi$.

A simple --- and admittedly {\sl ad hoc} --- way to build pairing into the dynamic friction function consists in adding to the r.h.s.\ of (\ref{eq:b3}) a slowly decaying term $\propto \,{\rm e}^{-t/\tau_{\rm esc}}$, corresponding to a contribution $\gamma_{\rm pairs}(\omega) = \gamma_0 / [\tau_{\rm esc}^{-1} - {\rm i}\omega]$ to $\gamma_\rho(\omega)$ and describing the random force correlations of pairs that unbind after a typical time $\tau_{\rm esc}$.
We interprete $\tau_{\rm esc}$ as the ``thermal escape time'' out of the logarithmic pair potential, screened at a distance $\xi$:
\begin{equation}
  \tau_{\rm esc}(\xi) \propto \frac{r_0^2 \Gamma}{q^2} \left( \frac{\xi}{r_0} \right)^z 
\ ,\quad
  z = \frac{q^2}{k_{\rm\scriptscriptstyle B} T}
\ .\label{eq:b11}\end{equation}
The dynamic exponent $z \approx 4.7$ in the vicinity of the KT transition.
Studying the correlations in an ensemble of {\em independent} pairs with average squared size $\langle {\scriptstyle \Delta} r^2 \rangle \approx 4b^2$, the amplitude of $\gamma_{\rm pairs}$ can be estimated as 
$\gamma_0 \approx q^4/k_{\rm\scriptscriptstyle B} T \, \langle {\scriptstyle \Delta} r^2 \rangle$.
The new term $\gamma_{\rm pairs}$ dominates $\gamma_\rho$ for large $\xi$ and low $\omega$, and so the static limit of the solution diverges as $\gamma_\rho(\omega\mbox{$=$}0) \propto \tau_{\rm esc}(\xi) \propto \xi^z$.
As a consequence, both the new characteristic frequency $\omega_{\rm ch,2} = \tau_{\rm esc}^{-1}$ and the static conductivity $\sigma(0) = q^2 \bar{n}/\gamma_\rho(0)$ of the plasma vanish smoothly as $\xi(T)^{-z}$.
The prediction $z \approx 4.7$ is in striking contrast to the scaling with $z=2$ in Fig.\ \ref{fig:scal}\,L obtained on the basis of AHNS theory, and to the usual expectation that $\sigma(0) \approx q^2 n_{\rm f}/\Gamma \sim \xi^{-2}$.
We note, however, that recent dynamical scaling analyses \cite{Pierson+99} for different realizations of 2D superfluids actually do suggest anomalously large dynamical exponents in the range $z \approx 4.5\,$...$\,5.9$.

\vspace{5mm}\noindent{\bf 4. CONCLUSIONS}\par\vspace{3mm}\noindent%
%
We have studied here the critical dynamics of a 2D, two-component plasma in the vicinity of the KT phase transition, using a modification of AHNS theory and a Mori type approach.
The AHNS approach is quite successful as a convenient phenomenological interpretation of simulation data, but conceptually dissatisfactory in that pairs are put in ``by hand''.
The Mori approach is more fundamental, starting from microscopic equations of motion, but the correlations it predicts (within a mode decoupling approximation) are not strong enough to imply critical slowing down.
We have proposed a way to cure this deficiency by introducing additional, long-lived random force correlations, describing temporary pairing.
A virtue of this extension is that it contains two frequency scales, one finite and the other vanishing at $T_{\rm\scriptscriptstyle KT}$, in agreement with the numerical findings.

\vspace{3mm}
\noindent
{\bf Acknowledgments:} The authors are grateful to Petter Minnhagen for helpful conversations.
D. B. thanks the organizers of the SCCS 99 for financial support.


\vspace{-3mm}
\begin{thebibliography}{99}
\parskip=-1.5mm


\bibitem{%
Minnhagen87}
P.~Minnhagen,
Rev.\ Mod.\ Phys.\ {\bf 59}, 1001--66 (1987).


\bibitem{%
Nelson83} 
D.~R.~Nelson, 
%
in {\em Phase Transitions and Critical Phenomena}, Vol.~7, 
edited by C.~Domb and J.~L.~Lebowitz, 
Academic Press, London, 1983, pp.~1--99.


\bibitem{%
Jonsson+Minnhagen97a}
A.~Jonsson, P.~Minnhagen, 
Phys.\ Rev.\ B {\bf 55}, 9035--46 (1997).

\bibitem{%
Theron+93} 
R.~Th\'eron, J.-B.~Simond, C.~Leemann, H.~Beck, P.~Martinoli, P.~Minnhagen, 
Phys.\ Rev.\ Lett.\ {\bf 71}, 1246--9 (1993).

\bibitem{%
anom_dyn} 
H.~Beck, 
Phys.\ Rev.\ B {\bf 49}, 6153--7 (1994);
S.~E.~Korshunov,
Phys.\ Rev.\ B {\bf 50}, 13616--21 (1994);
M.~Capezzali, H.~Beck, S.~R.~Shenoy, 
Phys.\ Rev.\ Lett.\ {\bf 78}, 523--6 (1997).




\bibitem{%
AHNS} 
V.~Ambegaokar, B.~I.~Halperin, D.~R.~Nelson, E.~D.~Siggia,
Phys.\ Rev.\ Lett.\ {\bf 40}, 783--6 (1978);
Phys.\ Rev.\ B {\bf 21}, 1806--26 (1980).



\bibitem{%
Hansen+McDonald86} 
J.-P.~Hansen, I.~R.~McDonald, 
{\em Theory of Simple Liquids} (2nd Edition), 
Academic Press Inc., San Diego, CA, 1986.

\bibitem{%
Pierson+99} 
S.~W.~Pierson, M.~Friesen, S.~M.~Ammirata, J.~C.~Hunnicutt, L.~A.~Gorham, 
Phys.\ Rev.\ B {\bf 60}, 1309--25 (1999).

\end{thebibliography}
\end{document}